\documentstyle[12pt]{article}
\begin{document}
\renewcommand{\thefootnote}{\fnsymbol{footnote}}
\setcounter{page}{1}
\begin{titlepage}
\begin{flushright}
\large
TSU HEPI 04-95 \\
November 1995
\end{flushright}
\vspace{3.cm}
\begin{center}
{\Large On the Flavour Dependence of the Mixed \\
 Quark-Gluon Condensate}
\vspace{0.5cm}

{\large Ketino Aladashvili\footnote
{E-mail: alada@bre.ge }
and Murman Margvelashvili\footnote{E-mail:
 mm@bre.ge, murman@cbdec1.cern.ch}}

\vspace{0.4cm}

{\em High Energy Physics Institute, Tbilisi State University \\
380086, Tbilisi, Georgia}

\vspace{0.3cm}

\end{center}
\vspace{1.0cm}
\begin{abstract}
The flavour dependence of the mixed quark-gluon condensate is studied
through the analysis of correlators of the hybrid
current $a_{\mu}=g\bar s\gamma_{\rho}\gamma_5 G_{\rho\mu}d$.
The flavour symmetry breaking for this type of condensates is found to be
less than that
for the quark condensates. For the ratio of strange to nonstrange condensates
we obtain \mbox{$R=0.95\pm 0.15$}. For the kaon coupling to the current
$a_\mu$ we find $\delta'^2=(0.020\pm0.005)GeV^2$, which is an order of
magnitude smaller than analogous chirally unsuppressed coupling.
\end{abstract}
\end{titlepage}
\renewcommand{\thefootnote}{\arabic{footnote}}
\setcounter{footnote}{0}
\newpage

\begin{section}*{Introduction}

The mixed quark gluon condensate $\langle0|\overline{s}
\sigma_{\mu\nu}G_{\mu\nu}s|0\rangle$ is an important phenomenological
parameter characterizing the structure of QCD vacuum. Considerable
attention has been paid in the past to determination of it's value, as
well as of the ratio

\begin{equation}
R= \frac{\langle\overline{s} \sigma_{\mu\nu}G_{\mu\nu}s\rangle }
{\langle \overline{d} \sigma_{\mu\nu} G_{\mu\nu}d\rangle}
\end{equation}
of strange and nonstrange quark-gluon condensates [1-5].

A QCD sum rule (SR) analysis of the baryon decuplet indicates the value
$R \simeq 1.3$
\cite{djn} while various hybrid sum rules favour the values in the lower
range $R=0.5-0.85$\cite{kkz,dosch,ovch}. At the same time, the analysis
of heavy to light mesons  \cite{er}  results in the restriction
$1-R<0.15$.

The two point functions of
the hybrid current $a_{\mu}=g\bar s\gamma_{\rho}\gamma_5 G_{\rho\mu}d$
which have been studied in \cite{kkz,dosch} by the QCD sum rules technique
are potentially the most sensitive to the value of $R$. The reason is,
that in this case the leading contributions come from the mixed
quark-gluon condensates of interest and the correlators are themselves
proportional to the $SU(3)$ breaking.
In the present work we continue the study of the correlators of the
hybrid current $a_\mu$ \cite{kkz,dosch}, being motivated by several
reasons:

First, we are interested in the value of the kaon coupling to the
hybrid current
$$\langle0|a_{\mu}|K(p)\rangle=-f_K\delta'^2p_{\mu}$$
which was left out in the the previous analyses.
This parameter measures the Chiral symmetry
breaking for kaon (in the chiral limit $\delta'^2=0$ \cite{nsvz}).
Besides, this quantity can contribute significantly to various kaon
transitions calculated in the Vertex Sum Rules approach (see e.g.
\cite{we1}).

Next, we have calculated some additional contributions to theoretical
expressions which happen to be important numerically and affect
the value of the strange quark-gluon condensate. The parameter set
which we use also differs from that of previous estimates.

For the determination of the unknown qantities we apply the recently
proposed procedure of simultaneous analysis of several sum rules
\cite{we2} which we believe provides the means of more detailed study
for the type of problems considered here. The allowed range of
unknown parameters is determined through the requirement of the
compatibility of different SRs which is tested using the
$\chi^2$ criterium.

\end{section}
\begin{section}*{Equations}
Consider the correlation functions of $a_\mu$ with pseudoscalar and
axial vector currents:
 \begin{eqnarray} ig \int dx e^{iqx}\langle0|T\left(\bar
d\gamma_{\mu}\gamma_5 s(x),\bar s\gamma_{\rho} \gamma_5 G_{\rho\nu}d(0)
\right)|0\rangle = g_{\mu\nu}\Pi^{ax}_1(q^2)+q_\mu q_\nu \Pi^{ax}_2(q^2)
\end{eqnarray}
\begin{eqnarray}
ig \int dx e^{iqx}\langle0|T\left(\bar d \gamma_5 s(x),\bar s\gamma_{\rho}
\gamma_5 G_{\rho\mu}d(0) \right)|0\rangle =  q_\mu  \Pi^{ps}(q^2)
\end{eqnarray}

\unitlength=0.90mm
\special{em:linewidth 0.4pt}
\linethickness{0.4pt}
\begin{picture}(140,90)(0,50)
\put(24.83,126.80){\oval(2.83,2.50)[rt]}
\put(27.71,126.93){\oval(2.92,2.42)[lb]}
\put(28.25,124.47){\oval(2.83,2.50)[rt]}
\put(31.12,124.60){\oval(2.92,2.42)[lb]}
\put(24.04,129.26){\oval(2.92,2.42)[lb]}
\put(23.92,128.05){\line(1,0){0.67}}
\put(24.50,128.05){\line(1,0){1.08}}
\put(24.67,128.05){\line(1,0){0.92}}
\put(25.58,128.05){\line(-1,0){0.33}}
\put(27.50,125.72){\line(1,0){1.17}}
\put(24.83,126.80){\oval(2.83,2.50)[rt]}
\put(32.30,122.14){\oval(2.83,2.50)[rt]}
\put(35.17,122.26){\oval(2.92,2.42)[lb]}
\put(35.71,119.80){\oval(2.83,2.50)[rt]}
\put(38.59,119.93){\oval(2.92,2.42)[lb]}
\put(31.33,123.39){\line(1,0){1.08}}
\put(34.96,121.05){\line(1,0){1.17}}
\put(31.00,123.39){\line(1,0){0.58}}
\put(38.16,118.72){\line(1,0){2.00}}
\put(26.17,121.16){\oval(34.33,18.33)[]}
\put(39.73,119.88){\oval(2.85,2.35)[rb]}
\put(42.47,119.90){\oval(2.62,1.79)[lt]}
\put(43.18,120.81){\circle*{1.54}}
\put(8.99,120.81){\circle*{1.54}}
\put(22.57,129.35){\line(0,1){0.89}}
\put(22.54,130.47){\circle*{0.94}}
\put(105.33,112.00){\line(1,0){34.67}}
\put(140.00,112.00){\line(0,1){22.67}}
\put(105.31,112.09){\line(0,1){22.67}}
\put(138.00,132.67){\line(1,1){4.00}}
\put(137.96,136.67){\line(1,-1){4.13}}
\put(134.40,115.63){\oval(2.92,2.42)[lb]}
\put(136.56,113.17){\oval(2.83,2.50)[rt]}
\put(135.59,114.42){\line(1,0){1.08}}
\put(139.57,113.40){\oval(2.92,2.42)[lb]}
\put(103.28,132.76){\line(1,1){4.00}}
\put(103.24,136.76){\line(1,-1){4.13}}
\put(105.33,112.14){\circle*{1.54}}
\put(139.85,112.14){\circle*{1.54}}
\put(105.21,129.30){\circle*{0.94}}
\put(56.67,52.34){\line(1,0){34.67}}
\put(91.34,52.34){\line(0,1){22.67}}
\put(56.65,52.43){\line(0,1){22.67}}
\put(89.34,73.01){\line(1,1){4.00}}
\put(89.29,77.01){\line(1,-1){4.13}}
\put(90.00,51.96){\oval(3.48,1.96)[rt]}
\put(54.61,73.10){\line(1,1){4.00}}
\put(54.57,77.10){\line(1,-1){4.13}}
\put(56.66,52.48){\circle*{1.54}}
\put(91.19,52.48){\circle*{1.54}}
\put(80.47,65.41){\oval(2.36,3.42)[rt]}
\put(82.89,65.35){\oval(2.47,3.54)[lb]}
\put(82.83,61.88){\oval(2.36,3.42)[rt]}
\put(85.25,61.82){\oval(2.47,3.54)[lb]}
\put(85.31,58.34){\oval(2.36,3.42)[rt]}
\put(87.72,58.28){\oval(2.47,3.54)[lb]}
\put(87.66,54.81){\oval(2.36,3.42)[rt]}
\put(90.08,54.75){\oval(2.47,3.54)[lb]}
\put(78.18,72.43){\oval(2.47,3.54)[lb]}
\put(78.23,68.95){\oval(2.36,3.42)[rt]}
\put(80.65,68.89){\oval(2.47,3.54)[lb]}
\put(76.94,71.95){\line(0,1){2.71}}
\put(74.96,72.77){\line(1,1){4.00}}
\put(74.92,76.77){\line(1,-1){4.13}}
\put(56.66,112.00){\line(1,0){34.67}}
\put(91.33,112.00){\line(0,1){22.67}}
\put(56.64,112.09){\line(0,1){22.67}}
\put(89.33,132.67){\line(1,1){4.00}}
\put(89.29,136.67){\line(1,-1){4.13}}
\put(79.18,119.08){\oval(2.83,2.50)[rt]}
\put(78.21,120.33){\line(1,0){1.08}}
\put(54.61,132.76){\line(1,1){4.00}}
\put(54.57,136.76){\line(1,-1){4.13}}
\put(56.66,112.14){\circle*{1.54}}
\put(91.18,112.14){\circle*{1.54}}
\put(78.37,119.04){\oval(3.44,2.58)[lt]}
\put(74.59,119.14){\oval(4.11,2.39)[rb]}
\put(74.30,117.94){\line(1,0){1.08}}
\put(74.45,116.66){\oval(3.44,2.58)[lt]}
\put(70.68,116.75){\oval(4.11,2.39)[rb]}
\put(69.90,115.55){\line(1,0){1.08}}
\put(70.06,114.27){\oval(3.44,2.58)[lt]}
\put(66.28,114.36){\oval(4.11,2.39)[rb]}
\put(66.62,111.88){\oval(3.44,2.58)[lt]}
\put(65.00,111.93){\circle*{0.94}}
\put(26.63,61.38){\oval(34.33,18.33)[]}
\put(43.64,61.03){\circle*{1.54}}
\put(9.45,61.03){\circle*{1.54}}
\put(44.37,61.82){\oval(4.48,1.77)[rb]}
\put(45.73,62.00){\oval(1.77,1.41)[rt]}
\put(45.49,65.30){\oval(2.47,1.65)[r]}
\put(45.55,63.59){\oval(2.36,1.77)[l]}
\put(45.49,68.72){\oval(2.47,1.65)[r]}
\put(45.55,67.01){\oval(2.36,1.77)[l]}
\put(45.49,72.14){\oval(2.47,1.65)[r]}
\put(45.55,70.43){\oval(2.36,1.77)[l]}
\put(27.34,73.32){\oval(2.47,1.65)[r]}
\put(27.40,71.61){\oval(2.36,1.77)[l]}
\put(27.28,75.85){\oval(2.36,3.42)[lb]}
\put(45.45,73.84){\oval(2.36,1.77)[l]}
\put(45.11,75.43){\oval(3.03,1.33)[rb]}
\put(26.15,76.57){\line(0,-1){1.04}}
\put(46.63,76.85){\line(0,-1){1.04}}
\put(27.59,70.60){\circle*{0.94}}
\put(24.12,74.18){\line(1,1){4.00}}
\put(24.08,78.18){\line(1,-1){4.13}}
\put(44.82,74.37){\line(1,1){4.00}}
\put(44.78,78.37){\line(1,-1){4.13}}
\put(81.93,119.14){\oval(2.92,2.42)[lb]}
\put(83.11,116.68){\oval(2.83,2.50)[rt]}
\put(82.14,117.93){\line(1,0){1.08}}
\put(81.81,117.93){\line(1,0){0.58}}
\put(85.95,117.00){\oval(2.92,2.42)[lb]}
\put(87.13,114.54){\oval(2.83,2.50)[rt]}
\put(86.16,115.79){\line(1,0){1.08}}
\put(85.83,115.79){\line(1,0){0.58}}
\put(89.97,114.77){\oval(2.92,2.42)[lb]}
\put(89.95,112.32){\oval(2.83,2.50)[rt]}
\put(134.39,114.43){\line(1,0){2.65}}
\put(135.38,114.43){\line(1,0){1.67}}
\put(129.40,118.18){\oval(2.92,2.42)[lb]}
\put(131.56,115.72){\oval(2.83,2.50)[rt]}
\put(130.59,116.97){\line(1,0){1.08}}
\put(129.39,116.98){\line(1,0){2.65}}
\put(130.37,116.98){\line(1,0){1.67}}
\put(124.39,120.54){\oval(2.92,2.42)[lb]}
\put(126.55,118.08){\oval(2.83,2.50)[rt]}
\put(125.58,119.33){\line(1,0){1.08}}
\put(124.39,119.33){\line(1,0){2.65}}
\put(125.37,119.33){\line(1,0){1.67}}
\put(119.39,123.09){\oval(2.92,2.42)[lb]}
\put(121.55,120.63){\oval(2.83,2.50)[rt]}
\put(120.58,121.88){\line(1,0){1.08}}
\put(119.39,121.88){\line(1,0){2.65}}
\put(120.37,121.88){\line(1,0){1.67}}
\put(114.39,125.73){\oval(2.92,2.42)[lb]}
\put(116.55,123.27){\oval(2.83,2.50)[rt]}
\put(115.58,124.52){\line(1,0){1.08}}
\put(114.38,124.53){\line(1,0){2.65}}
\put(109.39,128.28){\oval(2.92,2.42)[lb]}
\put(111.55,125.82){\oval(2.83,2.50)[rt]}
\put(110.58,127.07){\line(1,0){1.08}}
\put(109.38,127.08){\line(1,0){2.65}}
\put(110.36,127.08){\line(1,0){1.67}}
\put(106.54,128.18){\oval(2.83,2.50)[rt]}
\put(105.57,129.43){\line(1,0){1.08}}
\put(105.36,129.43){\line(1,0){1.67}}
\put(140.24,52.41){\line(0,1){22.67}}
\put(105.55,52.50){\line(0,1){22.67}}
\put(138.24,73.08){\line(1,1){4.00}}
\put(138.20,77.08){\line(1,-1){4.13}}
\put(103.52,73.17){\line(1,1){4.00}}
\put(103.48,77.17){\line(1,-1){4.13}}
\put(105.45,69.71){\circle*{0.94}}
\put(122.90,65.68){\oval(1.79,3.02)[b]}
\put(124.74,65.20){\oval(1.89,3.21)[t]}
\put(126.58,65.11){\oval(1.79,3.02)[b]}
\put(128.42,64.64){\oval(1.89,3.21)[t]}
\put(130.25,64.54){\oval(1.79,3.02)[b]}
\put(132.09,64.07){\oval(1.89,3.21)[t]}
\put(133.93,63.98){\oval(1.79,3.02)[b]}
\put(135.77,63.51){\oval(1.89,3.21)[t]}
\put(108.18,67.94){\oval(1.79,3.02)[b]}
\put(110.02,67.47){\oval(1.89,3.21)[t]}
\put(111.86,67.37){\oval(1.79,3.02)[b]}
\put(113.70,66.90){\oval(1.89,3.21)[t]}
\put(115.54,66.81){\oval(1.79,3.02)[b]}
\put(117.38,66.34){\oval(1.89,3.21)[t]}
\put(119.22,66.24){\oval(1.79,3.02)[b]}
\put(121.06,65.77){\oval(1.89,3.21)[t]}
\put(137.61,63.41){\oval(1.79,3.02)[b]}
\put(139.69,63.13){\oval(2.36,2.26)[lt]}
\put(105.87,67.66){\oval(2.83,4.15)[rt]}
\put(138.15,50.34){\line(1,1){4.00}}
\put(138.11,54.34){\line(1,-1){4.13}}
\put(103.62,50.34){\line(1,1){4.00}}
\put(103.58,54.34){\line(1,-1){4.13}}
\put(140.28,63.57){\circle*{1.54}}
\put(105.52,63.57){\circle*{1.54}}
\put(25.67,100.33){\makebox(0,0)[cc]{a)}}
\put(72.00,100.33){\makebox(0,0)[cc]{b)}}
\put(121.67,100.67){\makebox(0,0)[cc]{c)}}
\put(26.33,40.00){\makebox(0,0)[cc]{d)}}
\put(73.33,39.67){\makebox(0,0)[cc]{e)}}
\put(121.67,40.00){\makebox(0,0)[cc]{f)}}
\end{picture}
\vspace{1.00cm}
\begin{center}
{\bf Fig.1}
\end{center}

The operator product expansion for the invariant functions in (2), (3)
is obtained in the Euclidean region by calculating  diagrams shown in
fig.1 and has the following form:

\begin{eqnarray}
\Pi^{ps}(-Q^2)=
\frac{\langle \overline{d} \sigma G d\rangle-
\langle\overline{s} \sigma G s\rangle }{4Q^2}
+\frac{\alpha_s}{48\pi^3}Q^2\left(ln^2(Q^2/\mu^2)-ln(Q^2/\mu^2)\right)
 \nonumber \\
-\frac{\alpha_s}{3\pi}\left(\langle\overline{d}d\rangle-\langle\overline{s}s
\rangle\right)ln(Q^2/\mu^2) + \frac{m_s \langle \alpha_s /\pi G^{2}\rangle}
{8Q^{2}}\left( ln(Q^{2}/\mu^{2})-1\right)       \nonumber  \\
 + \frac{4\pi \alpha_s m_s}{27}
\frac{\left(3 \langle\overline{d}d\rangle^2 +\langle\overline{s}s\rangle^2
 -9 \langle\overline{d}d\rangle \langle\overline{s}s\rangle\right) \rho}
{Q^4}  \nonumber \\
-\frac{\pi^{2}}{9}\frac{\left(\langle\overline{d}d\rangle-
\langle\overline{s}s\rangle\right)\langle\frac{\alpha_s}{\pi}
G^{2}\rangle}{Q^4}
\end{eqnarray}

\begin{eqnarray}
\Pi_2^{ax}(-Q^2)=-\frac{m_{s}\langle\overline{s}\sigma G s\rangle}{6Q^{4}}
-\frac{2\alpha_s m_{s}}{9\pi Q^{2}}\left(5\langle\overline{s}s
\rangle-(ln(Q^2/\mu^2)+\frac{1}{3})
\langle\overline{d}d\rangle\right) \nonumber \\
+\frac{8\pi\alpha_s}{27}
\frac{\left( \langle\overline{s}s\rangle^{2}-
\langle\overline{d}d\rangle^{2}\right)\rho}{Q^4}
\end{eqnarray}

\begin{eqnarray}
\Pi_1^{ax}(-Q^2)=-\frac{ m_{s}\langle\overline{d}\sigma G d\rangle}{4Q^{2}}
+\frac{m_{s}\langle\overline{s}\sigma G s\rangle}{12Q^{2}} \nonumber \\
-\frac{\alpha_s}{3\pi}m_{s}\left(\langle\overline{s}s\rangle
-\frac{5}{3}\langle\overline{d}d\rangle\right)ln(Q^2/\mu^2)
+\frac{8\pi\alpha_s}{27}
\frac{\left( \langle\overline{s}s\rangle^{2}-\langle\overline{d}d
\rangle^{2}\right)\rho}{Q^{2}}
\end{eqnarray}

Here we work in the chiral limit $m_d=0$ for the $d$ quark and restrict
ourselves to the terms linear in $SU(3)$ breaking.
 The calculation has been performed in the $\overline{MS}$ scheme and
the fixed point gauge technique has been  used for condensate
contributions.

The quark and gluon condensate
contributions are given by the one loop diagrams b),c) and d). The
latter one requires the removal of infrared divergence which occurs due to
the mixing of the operators $m_sG^2$ and $ig\overline s G s$.
In spite of their one loop suppression, the quark and gluon condensate
contributions are quite important and
comparable to that of the quark-gluon condensates. Inclusion of
these terms shifts $R$ to its higher values, but mostly affects
$\delta'^2$ in comparison with \cite{dosch}.

In (4-6) the four quark condensates are reduced to the squares of
quark condensates, but we retain a factor $\rho$ to account
for the violation of vacuum dominance. The tree level dimension seven
contributions  have
been evaluated in the vacuum saturation approximation. However, due to
numerical smallness of these terms the approximation has  only minor
effect on our final results.

We keep the perturbative contribution (diagram a) )
 only for $\Pi^{ps}$ where it is of the order {\cal O}$(m_s)$ ,
for  $\Pi^{ax}_i$ the similar contribution is  proportional to
$m_s^2$ and thus negligible in our approximation.
The gluon condensate does not contribute to equations (5), (6)
to the considered order, since it is also {\cal O}$(m_s^2)$
\cite{dosch}.

The anomalous dimensions of the operators are accounted for by
taking e.g.

\begin{equation}
 \langle \bar q \sigma Gq\rangle (\mu) =
 \langle \bar q \sigma Gq\rangle(1GeV)\left(\frac{\alpha_s (1GeV)}
 {\alpha_s (\mu)}\right)^{31/54};~~
  m_s=m_s(1GeV)\left(\frac{\alpha_s(1GeV)}{\alpha_s(\mu)}\right)^{-4/9}
\end{equation}

We saturate the  phenomenological sides of SRs by the lowest lying
intermediate states:  the $K$ meson for the pseudoscalar correlator,
and the $K$ and $K_1$ for the axial vector one. We use the usual model
spectra with $K$ and $K_1$ as narrow resonances and the continuum, equal
to the theoretical one, starting at some threshold $s_0$. Following
\cite{dosch}, instead of the two nearby resonances
$K_1(1270)$ and $K_1(1400)$   we substitute an
effective resonance $K_1(1335)$.

After the Borel transformation \cite{svz} we obtain the following set of
equations:

\begin{eqnarray}\label{psb}
\frac{(1-R)\langle \overline{d} \sigma G d\rangle}{4M^2} +
\frac{\alpha_s}{3\pi}\left(\langle\overline{d}d\rangle
-\langle\overline{s}s\rangle\right)(1-e^{-x_0}) \nonumber \\
-\frac{\alpha_s}{48\pi^3}M^4\int^{x_0}_0e^{-x}x(1-2lnx)dx
+ \frac{m_{s}\langle\frac{\alpha_s}{\pi}
G^2\rangle}{8M^{2}}\left(\int^{x_0}_0\frac{1}{x}(1-e^{-x})dx-1\right)
\nonumber \\ + \frac{4\pi \alpha_s m_s}{27M^4}\rho
\left(3 \langle\overline{d}d\rangle^{2} +\langle\overline{s}s\rangle^{2}
- 9 \langle\overline{d}d\rangle \langle\overline{s}s\rangle \right)
- \frac{\pi^{2}}{9M^4}\left(\langle\overline{d}d\rangle-
\langle\overline{s}s\rangle\right)\langle\frac{\alpha_s}{\pi}
G^{2}\rangle = \nonumber \\
-\frac{f_K^2 m_K^2}{m_s M^2}
\delta^{\prime 2} e^{-m_K^2/M^2}
\end{eqnarray}

\begin{eqnarray}\label{axqb}
-\frac{m_{s}R\langle\overline{d}\sigma G d\rangle}{6M^{4}}
-\frac{2\alpha_s m_{s}}{9\pi M^{2}}\left(5\langle\overline{s}s
\rangle-\left( \int^{x_0}_0\frac{1}{x}(1-e^{-x})dx+\frac{1}{3}\right)
\langle\overline{d}d\rangle \right) \nonumber \\
+\frac{8\pi\alpha_s}{27M^4}\rho\left(
\langle\overline{s}s\rangle^{2}-\langle\overline{d}d\rangle^{2}\right)
=\frac{f_K^2\delta^{\prime2}}{M^2}e^{-m_K^2/M^2}
+\frac{C}{m_{K_1^2}M^2}e^{-m_{K_1}^2/M^2}
\end{eqnarray}

\begin{eqnarray}\label{axgb}
\frac{ m_{s}(R-3)\langle\overline{d}\sigma G d\rangle}{12M^{2}}
+\frac{\alpha_s}{3\pi}m_{s}\left(\langle\overline{s}s\rangle
-\frac{5}{3}\langle\overline{d}d\rangle\right)(1-e^{-x_0})
\nonumber \\
+\frac{8\pi\alpha_s}{27M^4}\rho\left(
\langle\overline{s}s\rangle^{2}-\langle\overline{d}d\rangle^{2}\right)
=- \frac{ C }{M^{2}} e^{-m_{K_1}^{2}/M^{2}}
\end{eqnarray}
where $C$ is the residue of the $K_1$ effective pole, and
$x_0=\frac{s_0}{M^2}$  with $s_0$ being the continuum onset.
The running parameters are taken at $\mu^2=M^2$.

Eqs.(8-10) constitute a system of sum rules to be used for the
determination of the unknown quantities $R$, $\delta'^2$ and $C$.
Following the previous analyses \cite{kkz,dosch} we could eliminate the
couplings $C$ and $\delta'^2$ from these equations and express $R$ as a
function of QCD parameters and $M^2$.  This however would mean that the
two of three equations are required to hold exactly at any $M^2$, while
the third one is considered as approximate equation with all uncertainties
and errors accumulated in it. Instead, we follow the procedure
developed in \cite{we2} and consider the Borel transformed SRs (8-10)
as a system of independent approximate equations, where we try to
estimate the expected errors and find a set of unknown parameters
which makes the whole system maximally consistent.

The  error analysis of such a system is the most
problematic task where the model assumptions are unavoidable.
Following \cite{we2} we model the errors of equations (8-10) in the
following way: For each one of these equations we choose some
reference point
$\tilde M^2_i$ (i=8,9,10) of the Borel parameter, where the
{\em relative} error $w_i$ is assumed to be minimal. Then the
 {\em absolute} error distribution is described by a simple function
\begin{eqnarray} \label{d}
      D(M^2)=\left\{ \begin{array}{ll}
        \frac{\tilde M^2-M_0^2}{M^2-M_0^2} & \mbox{if $M^2 <
	\tilde M^2~GeV^2$} \\
	1 & \mbox{if $M^2\geq\tilde M^2~GeV^2$}
	\end{array}
	\right.
\end{eqnarray}
which has a pole at $M_0^2<\tilde M^2$ (to account for the divergence
of the power series) and is constant for $M^2>\tilde M^2$.

Finally the {\em absolute} error is calculated as
\begin{eqnarray} \label{delt}
\Delta_i(M^2)=D(M^2)\omega_i A_i
\end{eqnarray}
where $A_i\equiv A_i(\tilde M^2)$ denote the r.h.s.'s of eqs.(8-10)
at the reference values of the Borel parameter. Note, that due to
the decrease of $A_i(M^2)$ at high $M^2$ eq.(\ref{delt}) corresponds
to growth of the {\em relative}  errors for both high and low values
of this parameter.

To get an idea of the scale of expected errors, we first examine
each of the equations (8-10) separately and
check their stability with respect to $M^2$ for different values
of  $\langle \overline sGs\rangle$. Fig.2 shows the $M^2$ dependence
of $\delta'^2$ and $C$ (normalized to their final values) as defined
from eqs.(8-10) with $R=0.85$. The  SR (\ref{psb}) is the
most sensitive to the value of $R$. It becomes more stable for higher
values of this parameter, but as can be seen from the figure, even in
this case the stability is not very good, which can be attributed to
the unknown higher orders corrections  or the roughness of the model
of the  phenomenological spectrum. On the other hand
 eq.(\ref{axqb}) is the most stable in $M^2$ for a wide range of
$R$ and we consider it to be the most reliable one.
For eq.(\ref{axgb}) the stability plateau starts at higher $M^2$ values
than for eqs.(9-10), which can be considered as indication of
the higher characteristic mass scale
for this equation $(m_{K_1}^2 \simeq 1.8GeV^2)$. Thus our check shows
that we should expect a greater relative error in (\ref{psb}) than in
other equations and use higher $M^2$ values for SR (10).
\end{section}

\begin{section}*{Results and Discussion}

In order to fix the unknown quantities $R$, $\delta'^2$ and $C$ we first
scan all their possible values. For each set we evaluate the equations
(8-10) at four separated $M^2$ points, within the
range of their expected validity.
Thus we build up a system of 12 linear approximate equations where for
each one we calculate the  errors according to eqs. (11), (12).
We look for a set of unknown parameters which makes these 12 equations
maximally consistent, e.i. gives minimal $\chi_{d.o.f}^2$. Once we
find such a solution, we test its
stability against variations of different parameters involved in
the derivation ($M^2$ points, $\omega_i$ etc.) and adjust the latter
to get the least sensitivite result.
Note, that this is not a standard $\chi^2$ procedure since in
our case the errors depend on the solutions themselves. The details and
justification of the approach can be found in \cite{we2}, where we have
used the similar method to restrict the values of standard condensates.

For eqs.(8),(9) we take the $M^2$ points to be in the range
$(0.8\div 2)~GeV^2$,
the effective pole in the error distribution is set to $M_0=0.7GeV^2$
and we use $\tilde M^2=1.2GeV^2$ as a reference value of $M^2$.
For eq.(11) we increase all these values by $(0.2 - 0.4)GeV^2.$
The typical relative errors which we have used in our
analysis are: $\omega_8=0.25$, $\omega_{9}=0.15$ and $\omega_{10}=0.20$.
This is quite a conservative choice, especially taking into account that
the relative errors grow for both higher and lower values of $M^2$.
The continuum onset in equations (9),(10) is taken as $s_0\simeq2.5GeV^2$,
while in the equation (\ref{psb}) where there is no $K_1$ contribution,
the continuum threshold is taken to be lower, $s_0\simeq1.5GeV^2$.

For the numeric evaluation we use the following values of the input
parameters:
\begin{eqnarray}
\Lambda_{MS}=150MeV, & m_s=180MeV \cite{jam,nar2},  \nonumber\\
\langle\bar qq\rangle= -(0.24\mbox{Gev})^3, &
\langle\bar ss\rangle=(0.7\pm0.1)\langle\bar dd\rangle \cite{nar1},
\end{eqnarray}
\vspace{-0.7cm}
$$\langle\frac{\alpha_s}{\pi} G^2\rangle= 1.2\cdot10^{-2}GeV^4 \cite{svz},
{}~~~\langle\overline dGd\rangle / \langle\overline dd\rangle =0.63GeV^2,
{}~~~ \rho=4\cite{we2}.\footnote{This corresponds to the results of
\cite{we2}where instead of the present value we have used
$\langle\sqrt{\alpha_s}\bar qq\rangle= -(0.24\mbox{Gev})^3$ of
\cite{svz}.}$$

The minimal $\chi^2$ solution is given by $R=0.93$,
$\delta'^2=0.021GeV^2$
and \mbox{$C=-6\cdot10^{-4}GeV^4$} corresponding  to $\chi^2_{min}=0.6$.
This solution is quite
stable with respect to the variations of different parameters introduced
in our procedure. The changes of $M^2$ points by $\pm 0.2GeV^2$,
of $w_i$s by 20\% and $M_0$ by $\pm0.1GeV^2$ do not affect the result
significantly. We have also checked the dependence on the parameters
of eq.(13).
So, changing ${\langle \overline s s\rangle}/{\langle \overline d
d\rangle}=0.6(0.8)$ we get $R=0.94(0.89)$, $\delta'^2=0.02(0.023)$;
$m_s=160(200)MeV$ gives $R=0.92(0.95)$, $\delta'^2=0.019(0.024)GeV^2$
and $M_0^2=0.8(0.5)GeV^2$ gives
$R=0.91(0.96)$ and $\delta'^2=0.024(0.019)GeV^2$ respectively.
The final result is also insensitive to
the variations of $s_0$ and $s_0'$.

In fig.3 we have plotted the ranges in the ($R, \delta'^2$) plane
corresponding to the solutions with
$\chi_{d.o.f}^2<1$ and $\chi_{d.o.f}^2<\chi^2_{min}+1=1.6$.
We consider all the solutions having $\chi_{d.o.f}^2<\chi^2_{min}+1=1.6$
as acceptable and thus quote our final result as
\begin{eqnarray}
R=0.95\pm 0.15~~~~~\delta'^2=(0.020\pm0.005)GeV^2
\end{eqnarray}
{}From the figure one can notice  that the correlation of these
quantities is insignificant.

The value of $R$ is larger than that in previous estimates [1-3] and is
consistent with the symmetry limit $R=1$. The errors are also bigger,
but to our feeling here they look more realistic.
The result conforms well with the
conjecture of \cite{er}, based on the analysis of heavy-to-light systems,
that inclusion of the gluon field into the operator reduces the strength
of the flavor symmetry violation.

The parameter $\delta'^2$ is an order of magnitude
smaller than the similar,
chirally unsuppressed coupling of the kaon to another hybrid current
$\bar s\gamma_{\nu}\tilde G_{\nu\mu}d$ ($\delta^2\sim 0.2GeV^2$
\cite{nsvz},\cite{ovch}). Consequently, it will not contribute
significantly to the
vertex sum rules except the cases proportional to the $SU(3)$
flavour symmetry breaking.

\end{section}

\begin{section}*{Acnowledgements}

We thank J.Gegelia for helpful discussions and V.Kartvelishvili for
carefully reading the manuscript.

One of us (M.M.) wants to express his gratitude to the Crystal Barrel
Collaboration and in particular to L.Montanet, N.Djaoshvili, M.Benayoun
and R.Ouared  for their kind hospitality at CERN where this work has
been finalized.
\end{section}

\newpage
\begin{center}
{\Large \bf Figure captions}
\end{center}

\vspace {3.cm}

{\bf Fig.2} Stability test for SRs (8-10) with $R=0.85$:

$\bullet$ --- $\delta'^2/(0.002GeV^2)$ from eq.(8);

$\times$ --- $\delta'^2/(0.002GeV^2)$ from eq.(10) with $C=6\cdot
10^{-4}GeV^4$;

+ ---  $C/(-6\cdot10^{-4}GeV^4)$ determined from eq.(9).

\vspace{2cm}

{\bf Fig.3}
The allowed region in the plane $(R,\delta'^2)$. The areas with
$\chi^2_{d.o.f}<1$ and \\ $\chi^2_{d.o.f}<\chi^2_{min}+1$ are indicated.

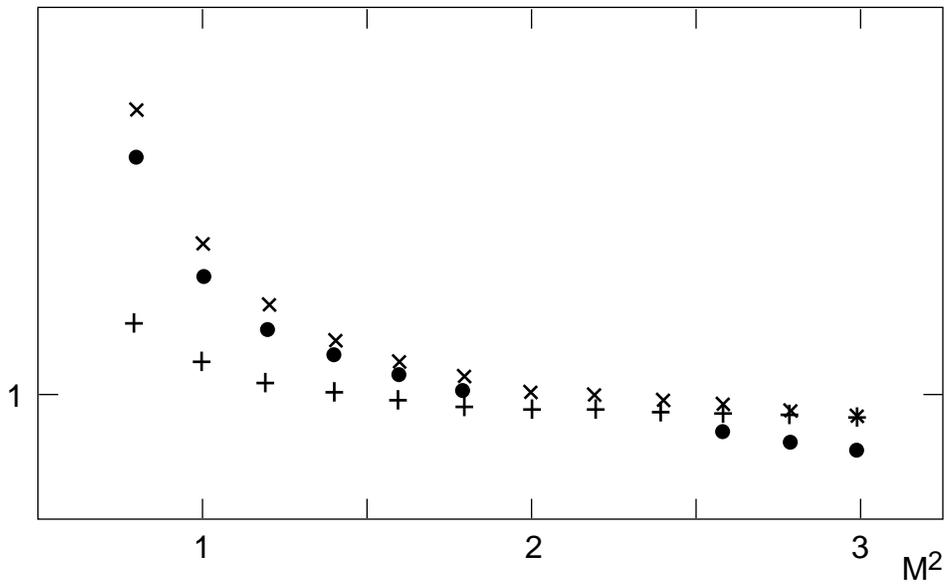
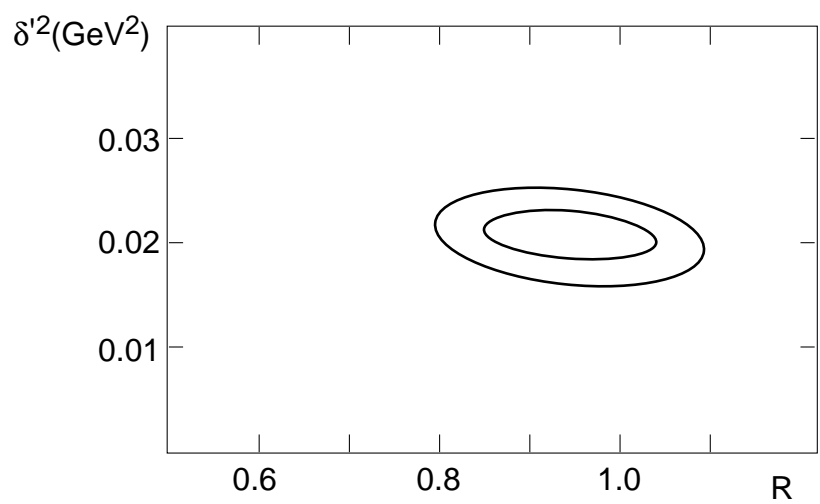
\end{document}